\documentclass[]{spie}  %>>> use for US letter paper
\usepackage{amsmath,amsfonts,amssymb}
\usepackage{mathrsfs} % ajoute des nouveau caractère math 

\usepackage{graphicx}
\usepackage{caption}
\usepackage{subcaption}
\graphicspath{{Images/}} %Setting the graphicspath
\usepackage[colorlinks=true, allcolors=blue]{hyperref}
\usepackage{relsize}

\usepackage{array}
\usepackage{booktabs}
\title{A Simulator-based Autoencoder \\ for Focal Plane Wavefront Sensing}

\author[a,b]{Maxime Quesnel}
\author[b]{Gilles Orban de Xivry}
\author[b]{Olivier Absil}
\author[a]{Gilles Louppe}
\affil[a]{Montefiore Institute of Electrical Engineering and Computer Science, University of Liège}
\affil[b]{Space sciences, Technologies and Astrophysics Research (STAR) Institute, University of Liège}

\authorinfo{Further author information: (Send correspondence to Maxime Quesnel)\\ E-mail: maxime.quesnel@uliege.be}

\begin{document} 
\maketitle

\begin{abstract}
Instrumental aberrations strongly limit high-contrast imaging of exoplanets, especially when they produce quasi-static speckles in the science images. With the help of recent advances in deep learning, we have developed in previous works an approach that applies convolutional neural networks (CNN) to estimate pupil-plane phase aberrations from point spread functions (PSF). In this work we take a step further by incorporating into the deep learning architecture the physical simulation of the optical propagation occurring inside the instrument. This is achieved with an autoencoder architecture, which uses a differentiable optical simulator as the decoder. Because this unsupervised learning approach reconstructs the PSFs, knowing the true phase is not needed to train the models, making it particularly promising for on-sky applications. We show that the performance of our method is almost identical to a standard CNN approach, and that the models are sufficiently stable in terms of training and robustness. We notably illustrate how we can benefit from the simulator-based autoencoder architecture by quickly fine-tuning the models on a single test image, achieving much better performance when the PSFs contain more noise and aberrations. These early results are very promising and future steps have been identified to apply the method on real data.
\end{abstract}


\section{INTRODUCTION}
\label{sec:intro} 
% FPWFS:
Imaging exoplanets requires instruments to have the capability of probing high contrasts between planets and their host star. Telescopes incorporating a coronagraph and adaptive optics (AO) tremendously help in this matter, but residual wavefront aberrations, in particular non-common path aberrations (NCPA) between the scientific and wavefront sensing arms, still greatly limit the detection of exoplanets\cite{Guyon:18}. Focal-plane wavefront sensing (FPWFS) methods\cite{Jovanovic:18} are particularly beneficial to measure NCPAs since the signature of these aberrations can be found in the focal-plane images.

% Autoencoder:
Deep learning with convolutional neural networks (CNN) has recently been implemented for focal-plane wavefront sensing\cite{Paine:18, Andersen:19, Andersen:20, Quesnel:20, Orban:21, Quesnel:22} with good results. CNNs offer great performance and robustness as well as fast inference, making them particularly adequate for NCPA correction during observing nights. One limitation of supervised learning using CNNs however is that the models require a ground truth to be trained. As building models with on-sky data is particularly challenging due to the lack of precise knowledge about the NCPAs themselves, it may be difficult to build reliable labelled datasets. Unsupervised learning approaches are thus particularly compelling, since they work without knowing the true phase aberrations. By reconstructing the input data, deep learning autoencoders are a well proven unsupervised approach. For focal-plane wavefront sensing the targets are the NCPAs instead of the point spread functions (PSF), and one way to extract this phase information with an autoencoder is to add physical knowledge to the model.

% Related works:
Several works in the field of optics have incorporated a physical simulator into their deep learning architecture. Bostan et al. 2020\cite{Bostan:20} and Wang et al. 2020\cite{Wang:20} for instance proposed each an unsupervised approach with an untrained neural network that reconstructs images on-the-fly based on an optical model. Peng et al. 2020 \cite{Peng:20} also incorporates an optical wave propagation model in the architecture to improve the quality of holographic images. Automatic differentiation of optical models has also been performed in the context of wavefront aberrations, notably for PSF reconstruction (Liaudat et al. 2022\cite{Liaudat:22}) as well as for phase retrieval and design (Wong et al. 2021\cite{Wong:21}).

% Contribution:
Motivated by the benefits a differentiable simulator can offer for an unsupervised approach, we define an autoencoder architecture that contains a CNN as the encoder and a optical simulator as the decoder. This allows to constrain the encoder to predict Zernike coefficients corresponding to the probed NCPAs. The method is described in Sec.~\ref{sec:method}. This unsupervised approach is tested on simulated data, and the results are presented in Sec.~\ref{sec:experiments}. We notably evaluate how the method handles photon noise and atmospheric turbulence residuals compared to using only a CNN. The possibility of fitting the models on an observed PSF that lies outside the training data distribution is also investigated. Developments for the approach to work with real instruments are finally discussed in Sec.~\ref{sec:conclusions}.

\section{SIMULATOR-BASED AUTOENCODER}
\label{sec:method}

\subsection{Data generation}
\label{sec:data_gen}

For the present study we consider a simple optical propagation between a pupil plane and a focal plane, so that the PSF is expressed as:
\begin{equation}
    PSF(x,y) = |\mathscr{F}[A(x,y)\,e^{\,i\,\Theta(x,y)}]|^2,
    \label{eq:phase_to_psf}
\end{equation}
where $A(x,y)$ is the pupil function and $\Theta(x,y)$ the pupil-plane phase aberrations. 

We construct the phase maps using Zernike polynomials following the Noll convention\cite{Noll:76}, starting from the tip mode, and based on an annular aperture with a central obstruction of 30\% of the total diameter. The sets of Zernike coefficients are randomly generated to approximate a $1/f^2$ power spectral density profile, and scaled to a given median root-mean-square (rms) wavefront error (WFE). In our simulations we consider two aberration regimes: NCPAs around 70 nm rms covering 20 Zernike modes, and 350 nm rms over 100 modes. The PSFs are computed using the \textsc{PROPER}\cite{Krist:07} optical propagation package. Our data is generated in the $K$ band at $\lambda = 2200$ nm, while we consider an aperture diameter of 10 m, a pixel scale of $\simeq 11.43$ mas/pix, and a field-of-view of $\simeq 1.47"$. The phase maps as well as the PSFs contain $64\times64$ pixels. In addition to the in-focus PSFs, we also generate out-of-focus PSFs with a defocus of $\lambda/4 = 550$ nm rms, which are used to lift the sign ambiguity that exists for the modes of even radial order, as typically done for focal-plane wavefront sensing\cite{Gonsalves:82, Vievard:19}. We add photon noise to the PSFs so that SNR $= \sqrt{N_{ph}}$, with $N_{ph}$ the number of photons. We also apply a square-root stretching operation to the PSFs, before normalizing the flux in the range [0,1]. This operation is done inside the encoder for the autoencoder architecture.  

For the experiments done in Sec.~\ref{sec:result_ao}, we also generate data containing atmospheric turbulence residuals. This is achieved by simulating an extreme AO system using the COMPASS library\cite{Ferreira:18}, giving a wavefront error of about 50~nm rms. To simulate a 1-s exposure in presence of a given amount of static NCPA, we use a sequence of 10 consecutive AO phase screens, and we sum up the corresponding PSFs. These AO simulations are described in more details in Quesnel et al. 2022\cite{Quesnel:22} (submitted).

\subsection{Autoencoder architecture}

\begin{figure}[t]
    \centering
    \includegraphics[scale=0.75]{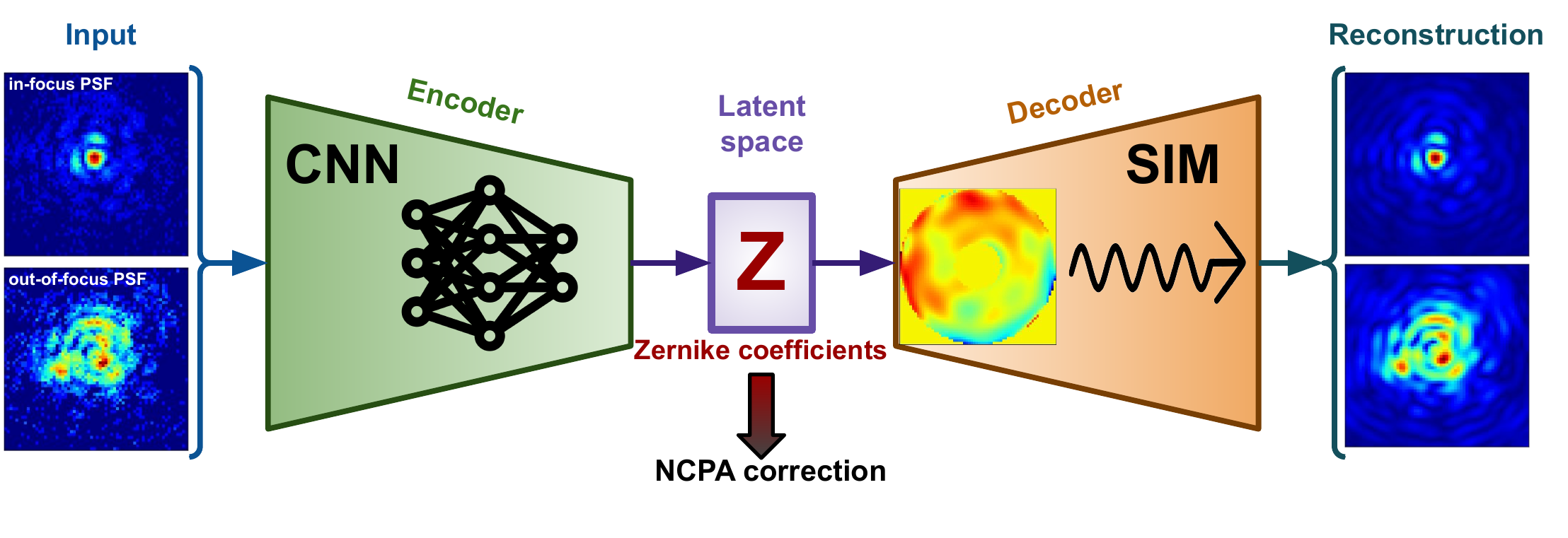}
    \caption{The proposed simulator-based autoencoder architecture (SimAE), using a CNN as the encoder and a differentiable optical simulator as the decoder. The latent space contains the Zernike coefficients which are used during training to reconstruct noiseless PSFs, while during inference these coefficients can be applied for NCPA correction.}
    \label{fig:archi}
\end{figure}

In this work we propose an autoencoder architecture, which is composed of an encoder and a decoder. The encoder is a CNN that maps Zernike coefficients from PSFs $x$, i.e., it approximates a non-linear function $f$ such that $z \approx f(x)$. We choose to use a state-of-the-art deep CNN called EfficientNet \cite{Tan:19}. This type of architecture stands out from other ones by using a new scaling technique, with all dimensions of the CNN (depth, width, and resolution) being scaled similarly. We choose to use the EfficientNet-B4 flavour, which offers the best trade-off between performance and runtime.

For our unsupervised approach, we also add a decoder which corresponds to a differential optical simulator, i.e. the decoder reconstructs the input PSFs based on the predicted Zernike coefficients, allowing to work without labels. This decoder is the same as the data generator defined in Eq.~\ref{eq:phase_to_psf}, with the exception that it produces noiseless PSFs. All the simulator's parameters are here fixed, while it may be beneficial in the future to consider learning parameters of the simulator for in-lab and on-sky applications. Regarding the latent space that exists between the encoder and the decoder, the number of Zernike modes has to be predefined. During inference on a trained model, the encoder containing the CNN allows to obtain the Zernike coefficients that correspond to the NCPAs of interest. Fig.~\ref{fig:archi} summarizes our simulator-based autoencoder architecture, which we call ``SimAE'' in the following sections. 

The loss function, which is used to optimize the neural network's weights in order to fit the model to the data, needs to be adapted to our SimAE architecture. Since the decoder reconstructs noiseless PSFs, the typical mean-square-error loss that makes the assumption of Gaussian noise is not appropriate on the PSF residuals. Since we are working on a simplified case where only photon noise is present in the images, i.e. each pixel follows a Poisson distribution, we can define our loss so that the log-likelihood of the Poisson probability distributions is maximized, giving the expression: 

\begin{equation}
    \mathcal{L}_{SimAE}(x;\phi) = - \mathbb{E}_{x \sim p(x)} \left[ \log \left( \frac{\lambda(x; \phi)^x}{x!}\exp(-\lambda(x; \phi)) \right)\right],
    \label{eq:loss_simae}
\end{equation}
where $\lambda(x; \phi)$ are the rates of the Poisson distributions, which depend on both the input PSFs $x$ sampled from the dataset distribution $p(x)$, and the parameters $\phi$ of the encoder (CNN). These rates correspond to the noiseless PSFs that are constructed by the decoder.

The baseline on which we compare our SimAE models is a standard CNN architecture. The loss function is in this case defined as the root-mean-square error of the phase residuals. It can be expressed using the Zernike coefficients as:

\begin{equation}
  \mathcal{L}_{CNN}(z, \widehat{z}(x; \phi)) = \sqrt{\frac{1}{N}{\sum_{i}^{N} (z_i - \widehat{z_i}(x; \phi))^2}},
\label{eq:loss_cnn}
\end{equation}
where $z$ and $\widehat{z}$ correspond to the true and predicted Zernike coefficients respectively, while $N$ is the number of Zernike modes (20 or 100). 

\section{EXPERIMENTS}
\label{sec:experiments}

\subsection{Protocol}

\begin{figure}[t]
    \centering
    \begin{subfigure}{.49\linewidth}
        \includegraphics[scale=0.25]{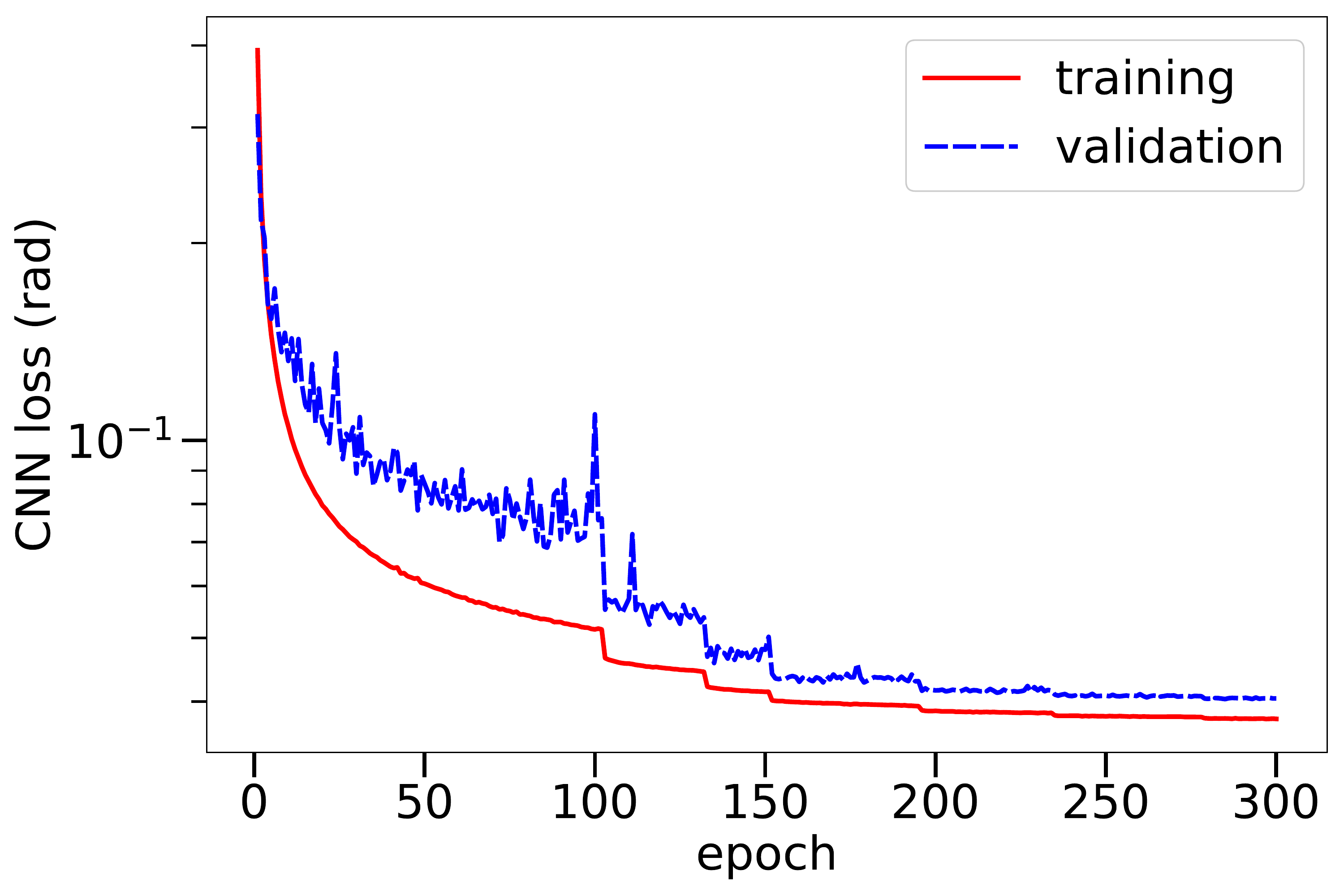}
    \end{subfigure}
    \begin{subfigure}{.49\linewidth}
        \includegraphics[scale=0.25]{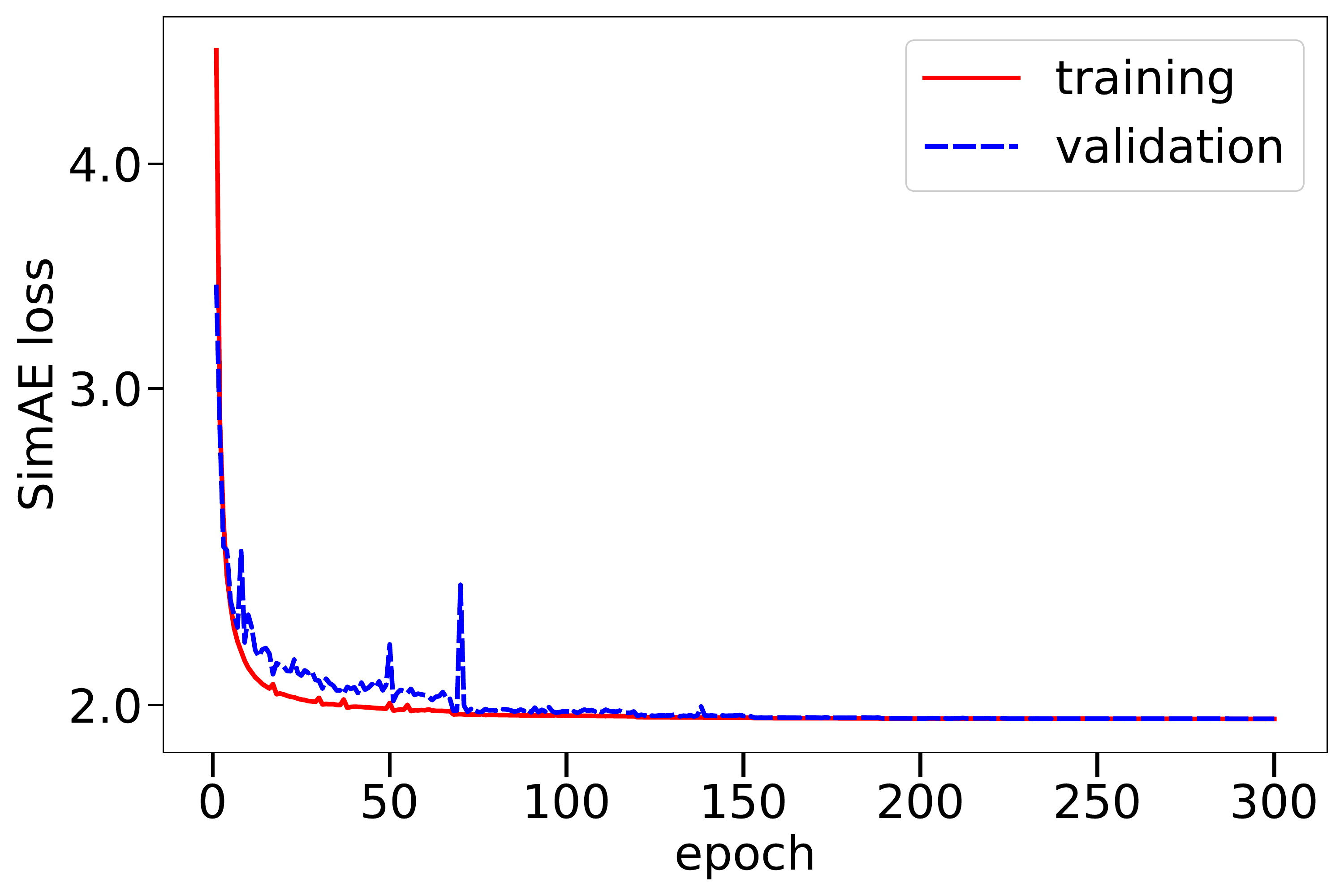}
    \end{subfigure}
    \caption{Examples of training and validation losses over the epochs for each approach: CNN (\textbf{left}) and SimAE (\textbf{right}). In these examples the models are trained on datasets containing 350 nm rms of aberrations distributed over 100 modes, and a SNR equal to 300.}
    \label{fig:loss_epoch}
\end{figure}

For a given training, a dataset composed of $10^{5}$ PSF pairs is randomly split into training (90\%) and validation (10\%) sets, with a batch size of 64. For the supervised learning approach with a CNN, each sample also contains the true NCPA phase maps as labels. The Adam optimizer \textbf{\cite{Kingma:17}} with a starting learning rate of $10^{-3}$ is used to update the weights of the CNN. The learning rate is decreased by a factor two as soon as the validation loss reaches a plateau over 15 epochs to improve the performance. Efficientnet-b4 weights are initialized with models pretrained on ImageNet. An example of the resulting training and validation losses for each method can be found in Fig.~\ref{fig:loss_epoch}. The models are then systematically evaluated on 1000 new samples.

\subsection{Results}

\subsubsection{Performance compared to a standard deep CNN}

We first compare the performance regarding each Zernike mode. The metric used is the root-mean-square error (RMSE) per mode, computed over the entire test set: 
\begin{equation}
    \sigma_{z} = \sqrt{\frac{1}{N_{\rm test}} \, \sum_{i}^{N_{\rm test}} (\hat{c}_i - c_i)^2} \:,
\end{equation}
where $N_{\rm test}$ is the number of test samples, while $\hat{c}$ and $c$ are the estimated and true Zernike coefficients, respectively.

\begin{figure}[t]
    \centering
    \includegraphics[scale=0.35]{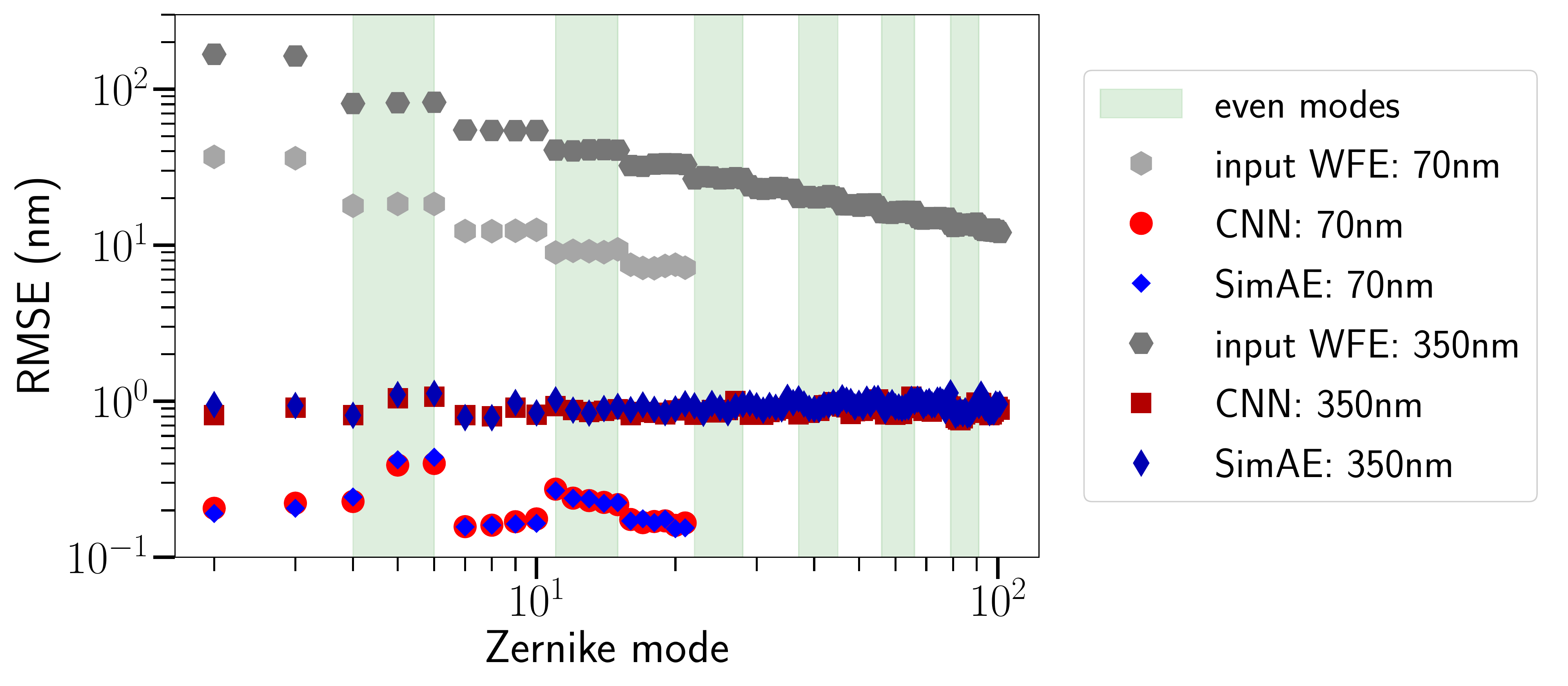}
    \caption{Performance per Zernike mode, for a low aberration regime (70 nm over 20 modes, light shades) and a high aberration regime (350 nm over 100 modes, darker shades), at a SNR of 1000. The CNN and SimAE architectures are compared (blue and red, respectively).}
    \label{fig:rmse_zern}
\end{figure}

As shown in Fig.~\ref{fig:rmse_zern}, the Zernike coefficients are very well reconstructed, even at high aberrations contents (350 nm rms over 100 modes). There is no particular distinction between even and odd modes, and the SimAE and CNN manifest very similar performance.

Since the SimAE is defined to take into account photon noise through its loss function, it is worth assessing the performance for different SNR levels. We thus compare the residual errors in terms of rms WFE on the total phase residuals. The metric is therefore defined for each test sample as:
\begin{equation}
    %\begin{split}
    \sigma_{\phi} = \sqrt{\frac{1}{N_{\rm pix}}\sum_{i}^{N_{\rm pix}} (\hat{\phi}_i - \phi_i)^2},
\end{equation}
where $N_{\rm pix}$ is the number of pixels within the pupil area, while $\hat{\phi}$ and $\phi$ are the estimated and true pupil phases, respectively. 

\begin{figure}[t]
    \centering
    \includegraphics[scale=0.35]{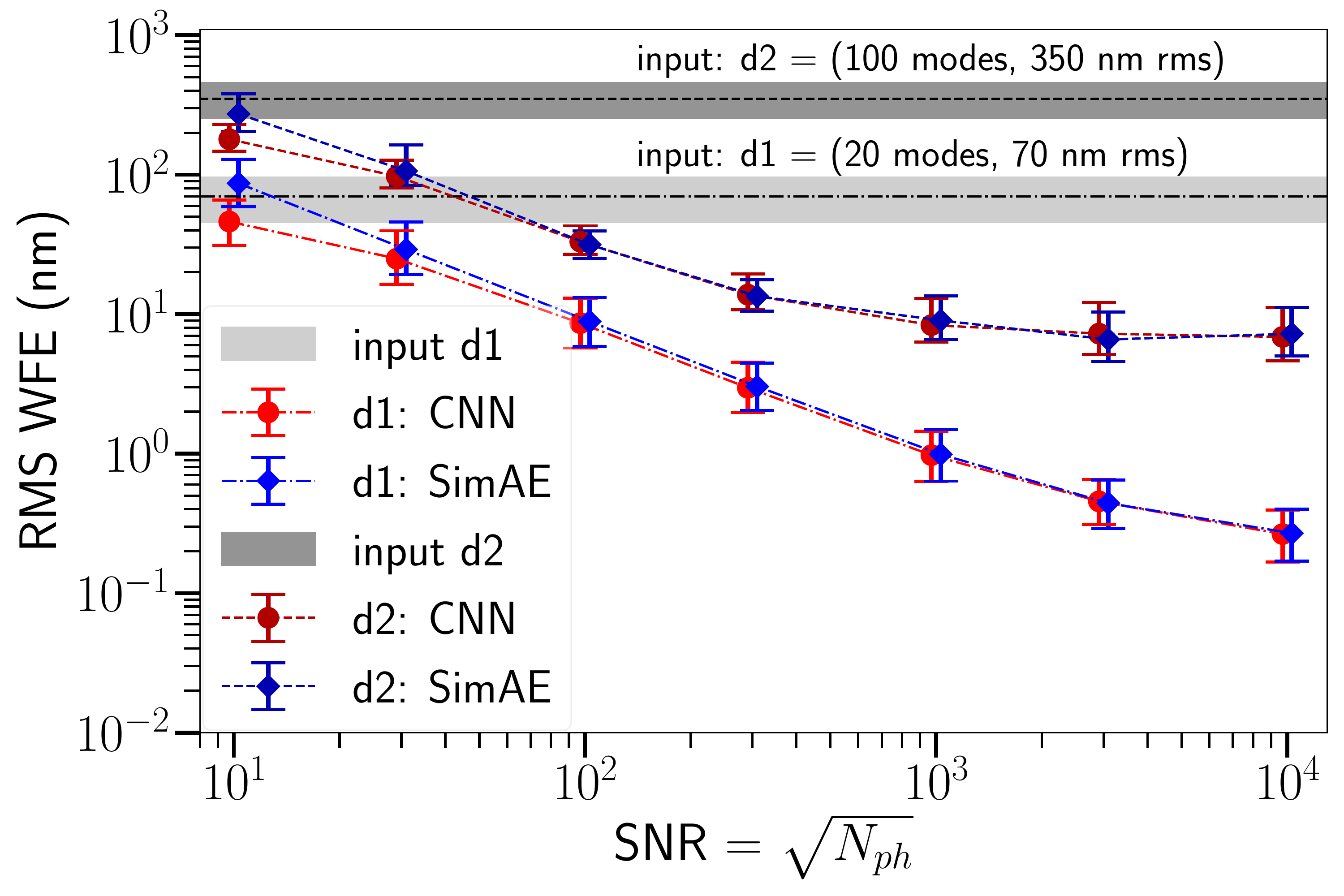}
    \caption{Performance at different SNR levels for the SimAE (blue) and the standard CNN (red) models, at low and high aberration regimes (darker and brighter shades respectively). The points correspond to the median value and the error bars to the 2-98th percentiles.}
    \label{fig:rmse_snr}
\end{figure}

The errors on the phase retrieval for a range of SNR levels are shown in Fig.~\ref{fig:rmse_snr}, where each point corresponds to a model trained and evaluated at the same given SNR. The two approaches show almost identical performance, and a plateau is reached around a SNR of 3000 in the high aberration regime. This has been observed and discussed in previous works\cite{Orban:21, Quesnel:22} as well. Only at very low SNRs the SimAE appear to differ from the CNN approach. This difference could be due to the loss function of the SimAE that is defined on the PSFs and not on the NCPAs, so that there is not the implicit regularization occurring with supersized CNNs, which has been described by Orban de Xivry et al. 2021\cite{Orban:21}. The residual errors thus tend to follow the theoretical limit for lower SNRs with the SimAE.

\subsubsection{Robustness of the trained models}
Deep learning models may show limitations when evaluated on data lying outside the training data distribution. It is therefore important to assess the stability of the models in such conditions, which would probably occur when applying the algorithm on sky.

\begin{figure}[t]
    \centering
    \begin{subfigure}{.49\linewidth}
        \includegraphics[scale=0.205]{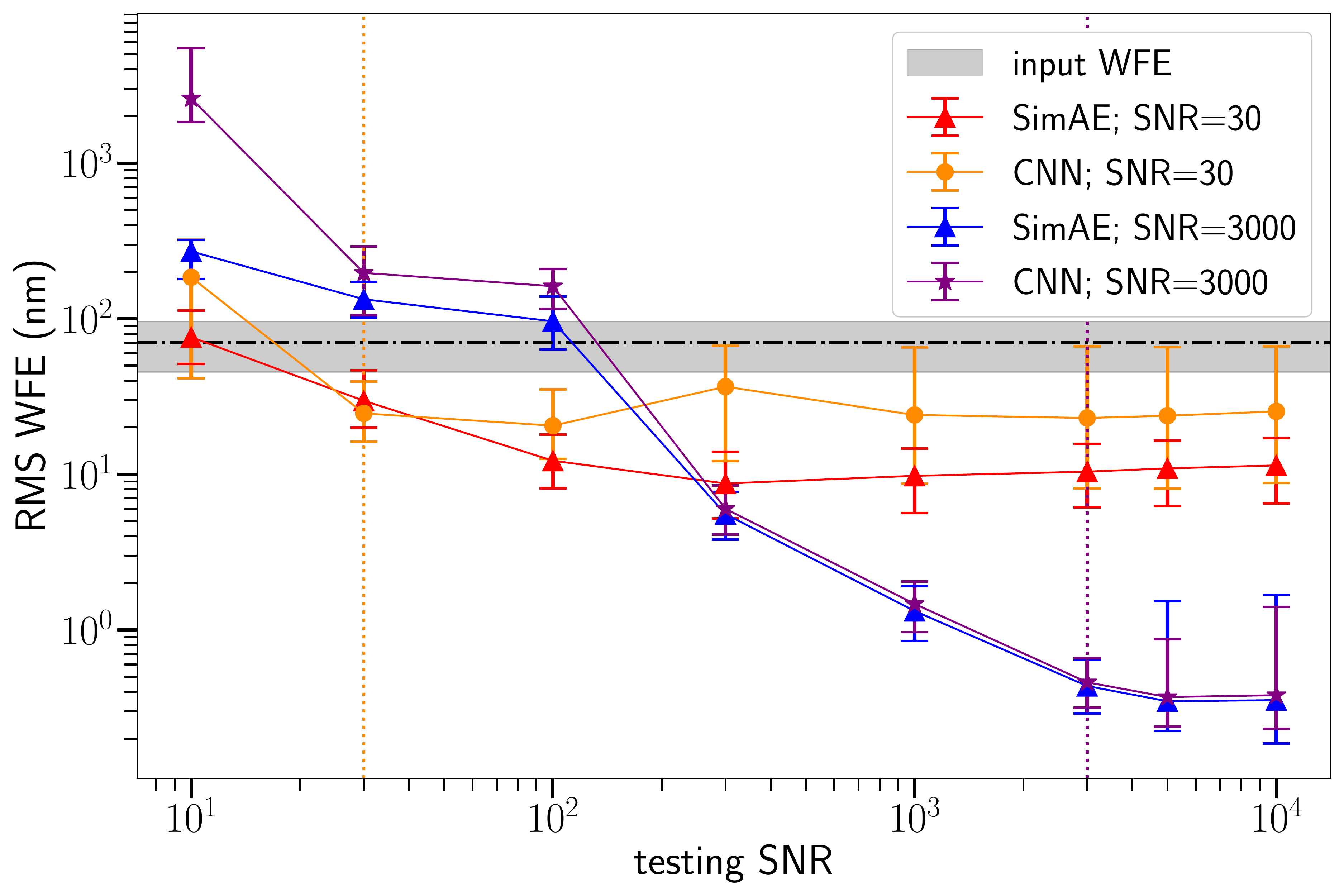}
    \end{subfigure}
    \begin{subfigure}{.49\linewidth}
        \includegraphics[scale=0.205]{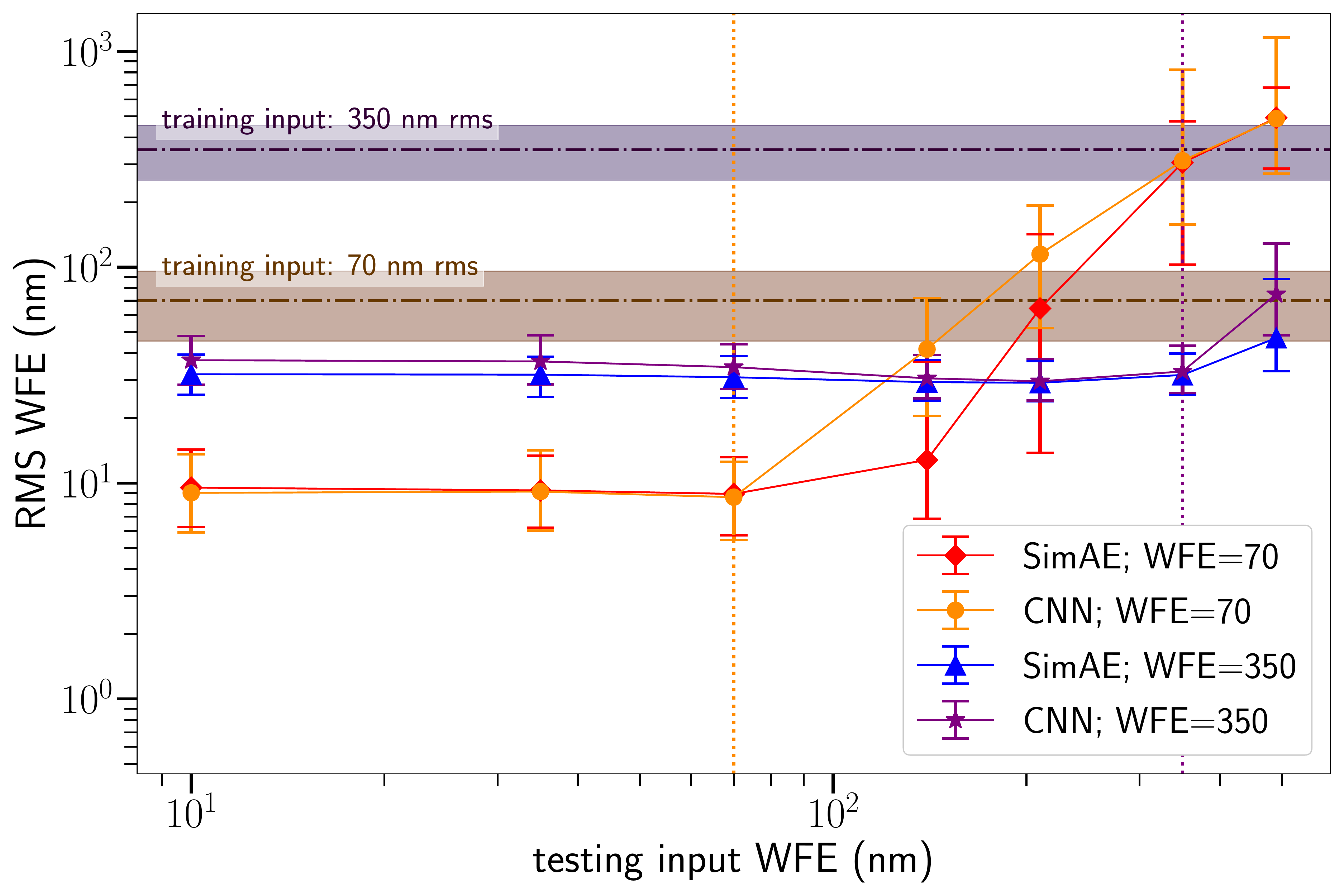}
    \end{subfigure}
    \caption{\textbf{Left:} Performance with changing the SNR of the test data for two models trained on SNRs of 30 and 300 (indicated by the vertical dotted lines). The median input WFE is 70 nm with 20 modes. \textbf{Right:} Performance for different testing input WFE, on models trained at 70 nm and 350 nms (20 and 100 modes respectively).}
    \label{fig:robust}
\end{figure}

Here we compare again the SimAE method with the CNN, this time varying the testing SNR (Fig.~\ref{fig:robust}, left), on a model trained at a median WFE of 70 nm rms distributed over 20 modes. Giving data containing less noise to the models always provides stable predictions, and we even obtain a marginal improvement with the SimAE when trained at a SNR of 30. If we introduce more noise to the test data, the error quickly increases, at about the same rate with both approaches. We also experiment by changing the testing input WFE in the data (Fig.~\ref{fig:robust}, right). Here the CNN and SimAE showcase identical behaviour, with constant performance for lower testing WFE than the training one, and strong degradation for higher testing WFE, although with the SimAE the performance does not degrade as quickly as with the CNN. Similar trends regarding the CNN has been found in previous works\cite{Orban:21, Quesnel:22}. Overall the SimAE method offers only slight improvement in terms of robustness compared to the CNN. To benefit the most from such a physics-based approach, it would be useful to include more knowledge about the data in the models, which could be provided by a variational autoencoder\cite{Kingma:13} approach for instance. 

\subsubsection{Transfer learning}

\begin{figure}[t]
    \centering
    \begin{subfigure}{.49\linewidth}
        \includegraphics[scale=0.25]{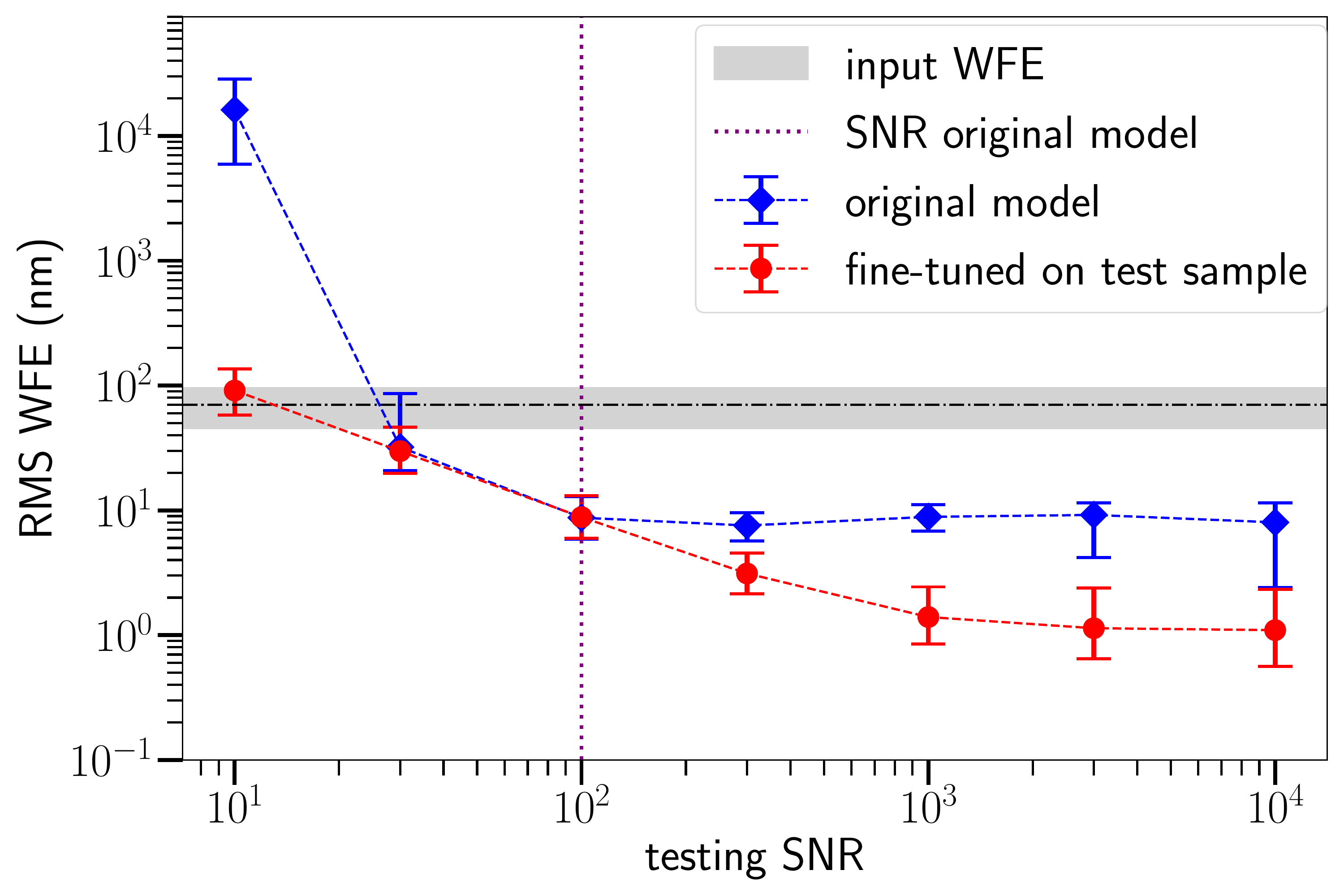}
    \end{subfigure}
    \begin{subfigure}{.49\linewidth}
        \includegraphics[scale=0.25]{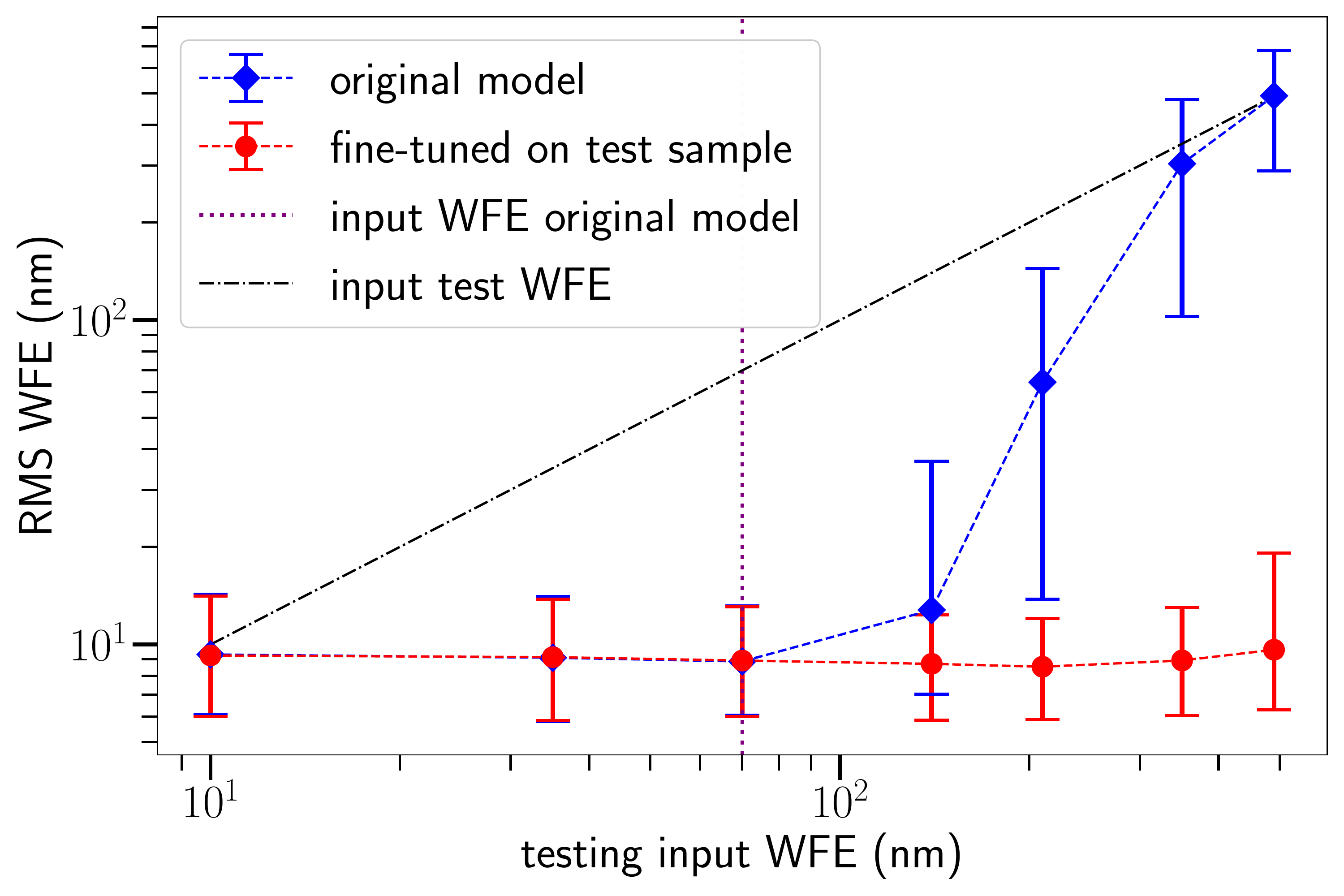}
    \end{subfigure}
    \caption{Performance of models fine-tuned on each test sample individually, for varying SNR (\textbf{left}) and input WFE levels (\textbf{right}). The model used for both plots was initially trained on data with 70 nm rms, 20 modes, and a SNR of 100, as indicated by the vertical dotted lines.}
    \label{fig:trans_learn}
\end{figure}

Instead of directly applying a trained model on observed data, we propose to fine-tune the model using a single ``on-sky'' image, initializing with the trained weights of the original model. This is done to illustrate how we could quickly adapt the model from simulated to on-sky data, which may slightly differ from the training data. To test this on our SimAE method, we use a model trained on data containing 70 nm rms and a SNR of 100. The levels of NCPA and SNR are changed on the test data, and the results are shown in Fig.~\ref{fig:trans_learn}. The models perform much better when transfer learning is done on each test sample, in particular when their SNR and WFE are higher than for the original training data. This second training on a single test sample is performed over 200 epochs, which takes about 20 seconds to complete on one GPU (Nvidia GeForce RTX 2080 Ti). For data relatively close to the ones used for pre-training, convergence is naturally achieved after much less epochs and the final prediction can be made in only a few seconds.   

\subsubsection{Atmospheric turbulence residuals}
\label{sec:result_ao}

\begin{figure}[t]
    \centering
    \includegraphics[scale=0.3]{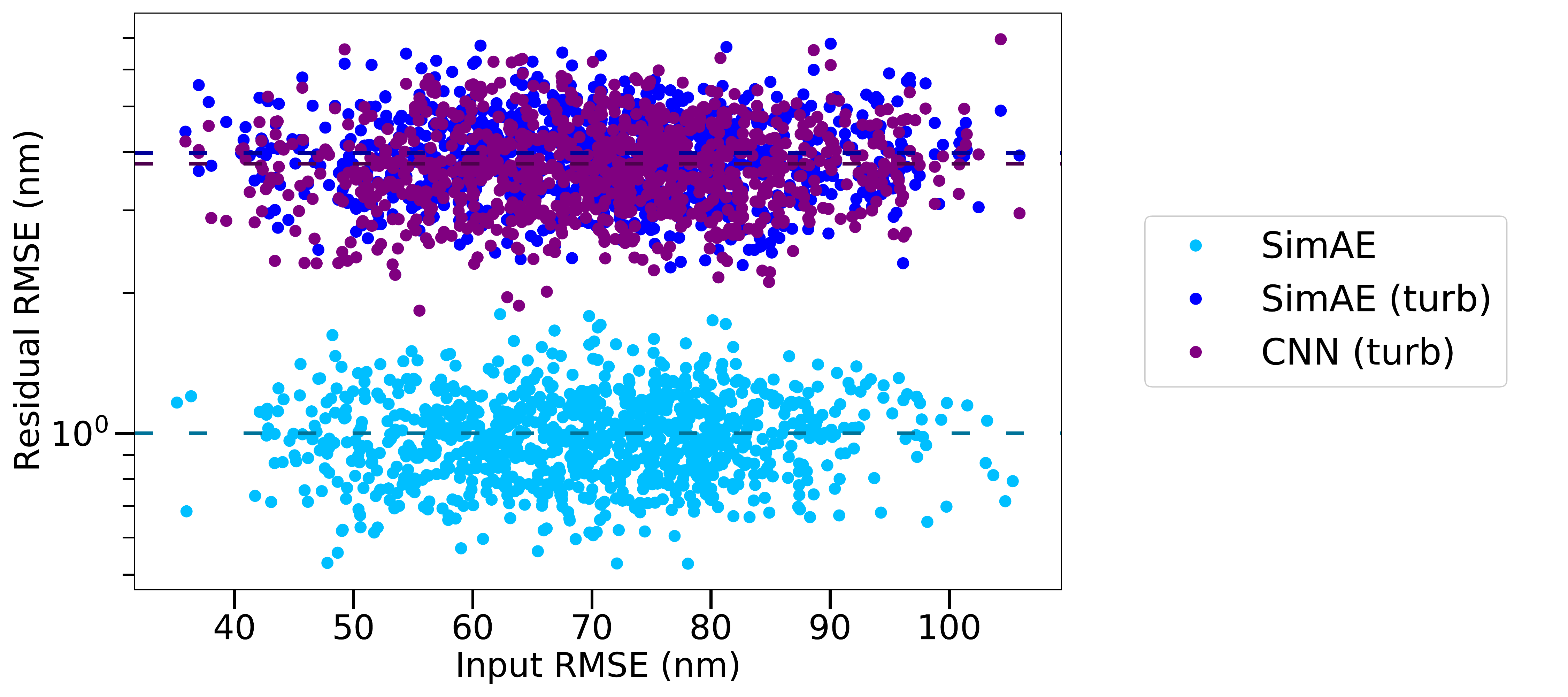}
    \caption{Residual errors as a function of the input WFE, for a model trained on a dataset with WFEs distributed around 70 nm (20 modes). A SimAE model with data containing only photon noise (cyan), is compared to CNN and SimAE models (purple and blue, respectively) trained on data containing also AO residuals, which are not provided to the decoder during training and evaluation.}
    \label{fig:rmse_turb}
\end{figure}

In the previous experiments, the decoder, i.e the differentiable optical simulator, was able to exactly reconstruct what our data generator produced to build our training and test sets, except for photon noise, but for which the loss function was adapted. In order to test the method on data the decoder cannot totally reconstruct, we introduce atmospheric turbulence residuals to the input data, without giving this information to the decoder. The way these AO residuals are added is described in Sec.~\ref{sec:data_gen}. Even with the presence of this additional source of noise, the training is very stable, and the evaluation results shown in Fig.~\ref{fig:rmse_turb} indicate that the SimAE suffers a reasonable degradation in performance compared to a case without turbulence. Using a CNN with labels that does not account for these AO residuals gives the same results.

\section{CONCLUSIONS AND FUTURE WORK}
\label{sec:conclusions}

We have developed an unsupervised deep learning method for focal-plane wavefront sensing by incorporating an optical simulator into a CNN-based autoencoder architecture (SimAE). We observe very similar performance compared to using a simple CNN, both approaches reaching the expected theoretical limit over a large range of conditions. The SimAE however has the major advantage of not requiring labels during the training of the models. One promising opportunity with an approach like this SimAE is that we can quickly fit the model to the currently observed PSF using a pre-trained model. We have shown in this paper that such fine-tuning of the models can greatly improve the results for data showing different noise or NCPA levels. Dealing with data the decoder cannot reproduce is naturally a potential limiting factor for the method. We have trained models with atmospheric turbulence residuals present on top of NCPAs, and the loss of performance that we obtain stays well constrained. In practice these turbulence residuals may even be given to the decoder, by using the wavefront sensor telemetry to learn what the AO residuals effectively are, with the help of a small neural network for example.

Several components of our architecture can be improved in the future. We may first benefit from variational inference, i.e. predicting posterior distribution for Zernike coefficients as well as to incorporate prior knowledge about them. This could yield better robustness for the models thanks to the additional information given during training and inference. Regarding the decoder, a more complete simulator, e.g. including a coronagraph for high-contrast imaging and other relevant instrumental components, will be essential to envision applying the technique on-sky with a real instrument. 

\acknowledgments % equivalent to \section*{ACKNOWLEDGMENTS}
For our study we have used the optical propagation package PROPER\cite{Krist:07} to generate our datasets. We have trained our deep learning models using the PyTorch\cite{Paszke:19} library while Sacred\cite{Greff:17} was used to manage our experiments.

The research was supported by the the Wallonia-Brussels Federation (grant for Concerted Research Actions), and by the European Research Council (ERC) under the European Union’s Horizon 2020 research and innovation program (grant agreement No 819155).

% References
\bibliography{main} % bibliography data in report.bib

\begin{thebibliography}{10}

\bibitem{Guyon:18}
{Guyon}, O., ``{Extreme Adaptive Optics},'' {\em Annual Review of Astronomy and
  Astrophysics}~{\bf 56},  315--355 (Sept. 2018).

\bibitem{Jovanovic:18}
{Jovanovic}, N., {Absil}, O., {Baudoz}, P., {Beaulieu}, M., {Bottom}, M.,
  {Cady}, E., {Carlomagno}, B., {Carlotti}, A., {Doelman}, D., {Fogarty}, K.,
  {Galicher}, R., {Guyon}, O., {Haffert}, S., {Huby}, E., {Jewell}, J.,
  {Keller}, C., {Kenworthy}, M.~A., {Knight}, J., {K{\"u}hn}, J., {Miller}, K.,
  {Mazoyer}, J., {N'Diaye}, M., {Por}, E., {Pueyo}, L., {Riggs}, A.~J.~E.,
  {Ruane}, G., {Sirbu}, D., {Snik}, F., {Wallace}, J.~K., {Wilby}, M., and
  {Ygouf}, M., ``{Review of high-contrast imaging systems for current and
  future ground-based and space-based telescopes: Part II. Common path
  wavefront sensing/control and coherent differential imaging},'' in [{\em
  Adaptive Optics Systems VI}{\nolinebreak\hspace{0.1em}]},  {Close}, L.~M.,
  {Schreiber}, L., and {Schmidt}, D., eds., {\em Society of Photo-Optical
  Instrumentation Engineers (SPIE) Conference Series} {\bf 10703},  107031U
  (July 2018).

\bibitem{Paine:18}
{Paine}, S.~W. and {Fienup}, J.~R., ``{Machine learning for improved
  image-based wavefront sensing},'' {\em Optics Letters}~{\bf 43},  1235 (Mar.
  2018).

\bibitem{Andersen:19}
{Andersen}, T., {Owner-Petersen}, M., and {Enmark}, A., ``{Neural networks for
  image-based wavefront sensing for astronomy},'' {\em Optics Letters}~{\bf
  44},  4618 (Sept. 2019).

\bibitem{Andersen:20}
{Andersen}, T., {Owner-Petersen}, M., and {Enmark}, A., ``{Image-based
  wavefront sensing for astronomy using neural networks},'' {\em Journal of
  Astronomical Telescopes, Instruments, and Systems}~{\bf 6},  034002 (July
  2020).

\bibitem{Quesnel:20}
{Quesnel}, M., {Orban de Xivry}, G., {Louppe}, G., and {Absil}, O., ``{Deep
  learning-based focal plane wavefront sensing for classical and coronagraphic
  imaging},'' in [{\em Society of Photo-Optical Instrumentation Engineers
  (SPIE) Conference Series}{\nolinebreak\hspace{0.1em}]},  {\em Society of
  Photo-Optical Instrumentation Engineers (SPIE) Conference Series} {\bf
  11448},  114481G (Dec. 2020).

\bibitem{Orban:21}
{Orban de Xivry}, G., {Quesnel}, M., {Vanberg}, P.~O., {Absil}, O., and
  {Louppe}, G., ``{Focal plane wavefront sensing using machine learning:
  performance of convolutional neural networks compared to fundamental
  limits},'' {\em Monthly Notices of the Royal Astronomical Society}~{\bf 505},
   5702--5713 (Aug. 2021).

\bibitem{Quesnel:22}
Quesnel, M., Orban~de Xivry, G., Absil, O., and Louppe, G., ``A deep learning
  approach for focal-plane wavefront sensing using vortex phase diversity,''
  (submitted, 2022).

\bibitem{Bostan:20}
Bostan, E., Heckel, R., Chen, M., Kellman, M., and Waller, L., ``Deep phase
  decoder: self-calibrating phase microscopy with an untrained deep neural
  network,'' {\em Optica}~{\bf 7},  559--562 (Jun 2020).

\bibitem{Wang:20}
Wang, F., Bian, Y., Haichao, W., Lyu, M., Pedrini, G., Osten, W., Barbastathis,
  G., and Situ, G., ``Phase imaging with an untrained neural network,'' {\em
  Light: Science \& Applications}~{\bf 9},  77 (05 2020).

\bibitem{Peng:20}
Peng, Y., Choi, S., Padmanaban, N., and Wetzstein, G., ``Neural holography with
  camera-in-the-loop training,'' {\em ACM Trans. Graph.}~{\bf 39} (nov 2020).

\bibitem{Liaudat:22}
Liaudat, T., Starck, J.-L., Kilbinger, M., and Frugier, P.-A., ``Rethinking
  data-driven point spread function modeling with a differentiable optical
  model,'' (2022).

\bibitem{Wong:21}
Wong, A., Pope, B., Desdoigts, L., Tuthill, P., Norris, B., and Betters, C.,
  ``Phase retrieval and design with automatic differentiation: tutorial,'' {\em
  Journal of the Optical Society of America B}~{\bf 38},  2465 (aug 2021).

\bibitem{Noll:76}
{Noll}, R.~J., ``{Zernike polynomials and atmospheric turbulence.},'' {\em
  Journal of the Optical Society of America (1917-1983)}~{\bf 66},  207--211
  (Mar. 1976).

\bibitem{Krist:07}
Krist, J.~E., ``{PROPER: an optical propagation library for IDL},'' in [{\em
  Optical Modeling and Performance Predictions
  III}{\nolinebreak\hspace{0.1em}]},  Kahan, M.~A., ed.,  {\bf 6675},  250 --
  258, International Society for Optics and Photonics, SPIE (2007).

\bibitem{Gonsalves:82}
Gonsalves, R.~A., ``{Phase Retrieval And Diversity In Adaptive Optics},'' {\em
  Optical Engineering}~{\bf 21}(5),  829 -- 832 (1982).

\bibitem{Vievard:19}
Vievard, S., Bos, S.~P., Cassaing, F., Ceau, A., Guyon, O., Jovanovic, N.,
  Keller, C., Lozi, J., Martinache, F., Mary, D., Montmerle-Bonnefois, A.,
  Mugnier, L., N\&apos;diaye, M., Norris, B., Sahoo, A., Sauvage, J.-F., Snik,
  F., Wilby, M.~J., and Wong, A., ``{Overview of focal plane wavefront sensors
  to correct for the Low Wind Effect on SUBARU/SCExAO},'' in [{\em {6th
  International Conference on Adaptive Optics for Extremely Large Telescopes,
  AO4ELT 2019}}{\nolinebreak\hspace{0.1em}]},  (June 2019).

\bibitem{Ferreira:18}
Ferreira, F., Gratadour, D., Sevin, A., and Doucet, N., ``Compass: An efficient
  gpu-based simulation software for adaptive optics systems,'' {\em 2018
  International Conference on High Performance Computing \& Simulation (HPCS)}
  ,  180--187 (2018).

\bibitem{Tan:19}
Tan, M. and Le, Q., ``{E}fficient{N}et: Rethinking model scaling for
  convolutional neural networks,'' in [{\em Proceedings of the 36th
  International Conference on Machine Learning}{\nolinebreak\hspace{0.1em}]},
  Chaudhuri, K. and Salakhutdinov, R., eds., {\em Proceedings of Machine
  Learning Research} {\bf 97},  6105--6114, PMLR (09--15 Jun 2019).

\bibitem{Kingma:17}
Kingma, D.~P. and Ba, J., ``Adam: A method for stochastic optimization,''
  (2017).

\bibitem{Kingma:13}
Kingma, D.~P. and Welling, M., ``Auto-encoding variational bayes,'' (2013).

\bibitem{Paszke:19}
Paszke, A., Gross, S., Massa, F., Lerer, A., Bradbury, J., Chanan, G., Killeen,
  T., Lin, Z., Gimelshein, N., Antiga, L., Desmaison, A., Kopf, A., Yang, E.,
  DeVito, Z., Raison, M., Tejani, A., Chilamkurthy, S., Steiner, B., Fang, L.,
  Bai, J., and Chintala, S., ``Pytorch: An imperative style, high-performance
  deep learning library,'' in [{\em Advances in Neural Information Processing
  Systems}{\nolinebreak\hspace{0.1em}]},  Wallach, H., Larochelle, H.,
  Beygelzimer, A., d\textquotesingle Alch\'{e}-Buc, F., Fox, E., and Garnett,
  R., eds.,  {\bf 32},  8026--8037, Curran Associates, Inc. (2019).

\bibitem{Greff:17}
{K}laus {G}reff, {A}aron {K}lein, {M}artin {C}hovanec, {F}rank {H}utter, and
  {J}\"urgen {S}chmidhuber, ``{T}he {S}acred {I}nfrastructure for
  {C}omputational {R}esearch,'' in [{\em {P}roceedings of the 16th {P}ython in
  {S}cience {C}onference}{\nolinebreak\hspace{0.1em}]},  {K}aty {H}uff, {D}avid
  {L}ippa, {D}illon {N}iederhut, and {P}acer, M., eds.,  49 -- 56 (2017).

\end{thebibliography}
\bibliographystyle{spiebib} % makes bibtex use spiebib.bst

\end{document}